\documentclass[aps,prc,twocolumn,floatfix,superscriptaddress,showpacs]{revtex4-1}
\usepackage{graphicx}
\usepackage[pdftex,colorlinks,citecolor=blue,bookmarks]{hyperref}
\usepackage{bm}
\usepackage{longtable}
\begin{document}
\title{
Occupation numbers of spherical orbits in self-consistent beyond-mean-field methods}

\author{Tom\'as R. Rodr\'iguez}
\affiliation{Departamento de F\'isica Te\'orica, Universidad Aut\'onoma de Madrid, E-28049 Madrid, Spain}
\author{Alfredo Poves}
\affiliation{Departamento de F\'isica Te\'orica, Universidad Aut\'onoma de Madrid, E-28049 Madrid, Spain}
\affiliation{Instituto de F\'isica Te\'orica, UAM-CSIC E-28049 Madrid,Spain}
\affiliation{USIAS fellow, Universit\'e de Strasbourg, France}
\author{Fr\'ed\'eric Nowacki}
\affiliation{Institute Pluridisciplinaire Hubert Curien, 23 rue du Loess, F-67037 Strasbourg Cedex 2, France}
\date{\today}
\begin{abstract}
We present a method to compute the number of particles occupying spherical single-particle (SSP) levels within the energy density functional (EDF) framework. These SSP levels are defined for each nucleus by performing self-consistent mean-field calculations. The nuclear many-body states, in which the occupation numbers are evaluated, are obtained with a symmetry conserving configuration mixing (SCCM) method based on the Gogny EDF. The method allows a closer comparison between EDF and shell model with configuration mixing in large valence spaces (SM-CI) results, and can serve as a guidance to define physically sound valence spaces for SM-CI calculations. As a first application of the method, we analyze the onset of deformation in neutron-rich $N=40$ isotones and the role of the SSP levels around this harmonic oscillator magic number, with particular emphasis in the structure of $^{64}$Cr. 
\end{abstract}
\maketitle
\section{Introduction}\label{Intro}
The nuclear shell model  (SM) is likely the most widely used framework to describe microscopically the structure of the atomic nucleus. In its simplest and naivest version, the nuclear many-body states are described in terms of products of one-body states, i.e.,  protons and neutrons occupy individual orbits defined by certain quantum numbers associated to the solution of the one-body problem in spherical coordinates -principal ($n$), orbital angular momentum ($l$) and total angular momentum ($j$) and its third component $(m_{j})$ quantum numbers. The idea behind this approach is the existence of an underlying spherical mean-field, made of a sum of one-body central and spin-orbit interactions that provide such orbits, and a residual interaction containing everything else. In the current versions of the shell model, the product-like mean-field states are used as the many-body basis where a diagonalization of the nuclear Hamiltonian is performed. 

In general, two different types of SM calculations can be distinguished depending on whether a core is considered in the definition of the system of interacting particles or not. In the more widely used large scale shell model with a core, Shell Model with Configuration Mixing in Large valence spaces (LSSM) or (SM-CI)~\cite{RMP_77_427_2005,PPNP_47_319_2001}, a valence space consisting of one or two major harmonic oscillator shell, sometimes different for protons and neutrons and sometimes with some added or removed orbits, is defined. In this case, single-particle levels below (above) the valence space are considered to be full (empty) and the nuclear interaction is renormalized to incorporate the relevant physics in such a reduced valence space. On the other hand, all particles are active in the no-core shell model (NCSM) approach~\cite{JPG_36_083101_2009} and the valence space must include many major harmonic oscillator shells in order to achieve convergence. However, since the number of states in the many-body basis increases combinatorially with the size of the valence space, the computational cost of either SM applications can become extremely large. Hence, the applicability of LSSM and NCSM is restricted to nuclei whose physical valence spaces produce m-scheme dimensions below 10$^{11}$~\cite{FN_PC} and to relatively light nuclei~\cite{PRL_107_072501_2011,PPNP_69_131_2013}, respectively.

Nuclear energy density functionals (EDF) based on Skyrme, Gogny or relativistic interactions can be applied in a more general manner along the whole nuclear chart~\cite{RMP_75_121_2003}. Initially designed to find the optimal self-consistent mean-field through Hartree-Fock (HF) or Hartree-Fock-Bogoliubov (HFB) calculations, these methods have been extended thoroughly in the last fifteen years to include beyond-mean-field (BMF) correlations needed to describe, for example, spectra of atomic nuclei. In particular, a more general form for the many-body states that considers linear combinations of different symmetry restored mean-field states has been implemented.

Because SM and EDF are the two main workhorses that provide a microscopic description of the structure of the nucleus, links between these frameworks are very useful. However, EDF methods tend to break most of the symmetries of the interaction at the mean-field level, i.e., they are defined in an intrinsic frame. Additionally, these are no-core calculations and the number of major harmonic oscillator shells included is generally much larger than in the LSSM and NCSM approaches. These aspects make SM and EDF states difficult to connect although several attempts have been already done. Notice however that, as dicussed at length in reference ~\cite{RMP_77_427_2005}, the SM-CI approach is implicitely based in the spherical mean field produced by an underlying, virtual, Hartree-Fock calculation.

For example, comparisons between SM calculations and EDF based on Gogny interactions~\cite{Berger84} were performed to describe the deformed nucleus $^{48}$Cr~\cite{PRL_75_2466_1995}, the triaxiality near $^{78}$Ni~\cite{PRC_88_034327_2013} or several aspects of neutrinoless double-beta decay nuclear matrix elements in the $pf$-shell~\cite{PRC_90_024311_2014}. Recently, the inclusion of cranked intrinsic states has proven an outstanding agreement between EDF and SM results for the excitation energies of magnesium isotopes and the nucleus $^{44}$S~\cite{PLB_746_341_2015,PRL_accepted}. Furthermore, SM valence spaces and interactions have been used to perform constrained HF calculations in exotic Ni isotopes and the SM states analyzed in terms of intrinsic quadrupole deformations~\cite{PRC_89_031301_2014}. Additionally, angular momentum projection before the variation method with a SM interaction has been also used to extract the intrinsic deformations of the nucleus $^{44}$S~\cite{PRL_114_032501_2015} and the structure of $sd$-shell nuclei~\cite{PRC_92_064310_2015}. 

On the other hand, EDF underlying interactions can be used to perform SM diagonalizations~\cite{PRC_78_024305_2008,PRC_85_044315_2012} and multipole decompositions~\cite{RMP_77_427_2005}. Furthermore, self-consistent mean-field analyses of single-particle energies in the deformed basis with Nilsson-like plots are routinely done to understand qualitatively the orbits that play a role for a given nucleus. In some cases, relevant deformed mean-field states have been studied in terms of the particle-hole structure in a spherical basis (see, for instance, Refs.~\cite{PRL_75_2466_1995,PRL_93_082502_2004}). However, a more quantitative analysis of the occupancies of spherical shells from EDF states including beyond-mean-field effects like symmetry restorations and configuration mixing is still missing.

In this paper a way to extract such occupancies is presented. The interest of these non directly observable quantities is two-fold. On the one hand, it allows a better comparison of the internal structure of the EDF states with SM states. On the other hand, the importance of each spherical orbit can be forecast and the method can serve as a guidance to define SM valence spaces, in particular, in mid-shell nuclei.

The paper is organized as follows. First, the EDF approach used here, the so-called symmetry conserving configuration mixing (SCCM) method, is reviewed (Sec.~\ref{SCCM}). Then, in Sec.~\ref{OCC_NUM} the spherical reference state and the formalism to compute spherical occupation numbers within the present EDF framework are discussed. A first application to analyze the role of the neutron $gds$ shell in $N=40$ neutron rich isotones is presented in Sec.~\ref{Results1}. Finally, the main conclusions are drawn in Sec.~\ref{Summary_outlook}. 
\section{Symmetry conserving configuration mixing method}\label{SCCM}
In this section the EDF method used in this work is summarized. A more detailed description with Skyrme, Gogny and relativistic interactions can be found in Refs.~\cite{RMP_75_121_2003,NPA_709_201_2002,PRL_99_062501_2007,PRC_78_024309_2008,PRC_81_064323_2010,PRC_81_044311_2010}. The starting point of the present SCCM method is the definition of the many-body states ($|\Phi^{JM;NZ;\sigma}\rangle$) as a linear combination of different symmetry restored HFB-like states within the generator coordinate method (GCM) framework~\cite{RingSchuck}:
\begin{equation}
|\Phi^{JM;NZ;\sigma}\rangle=\sum_{\vec{q}}\sum_{K=-J}^{J}f^{J\sigma}_{\vec{q},K}\hat{P}^{N}\hat{P}^{Z}\hat{P}^{J}_{MK}|\phi_{\vec{q}}\rangle
\label{GCM_general}
\end{equation}
where $J$, $M$, $N$ and $Z$ are the total and third component of the angular momentum and the number of neutrons and protons, respectively. Furthermore, $\sigma=1,2,3,...$ labels the different states for a given value of the angular momentum sorted from the lowest to the largest energies.

In the present application, particle number and angular momentum projections are performed using their respective projection operators $\hat{P}^{N}$, $\hat{P}^{Z}$ and $\hat{P}^{J}_{MK}$~\cite{RingSchuck}. Additionally, collective coordinates are also restricted to quadrupole deformations -$\vec{q}\equiv (\beta_{2},\gamma)$- although other degrees of freedom such as octupole deformation and parity projection~\cite{PRL_80_4398_1998,JPG_42_055109_2015}, pairing fluctuations~\cite{PLB_704_520_2011}, cranking frequencies~\cite{PLB_746_341_2015,PRL_accepted} and some others~\cite{PRC_71_044313_2005} have been successfully implemented, but require a much larger computational burden. 

Intrinsic HFB-like states, $|\phi_{\beta_{2},\gamma}\rangle\equiv|\rangle$, are found by minimizing the particle number projected HFB energy within the so-called variation after particle number projection method (PN-VAP)~\cite{NPA_696_467_2001,refff}, i.e.:
\begin{equation}
E'_{\beta_{2},\gamma}=\frac{\langle\hat{H}\hat{P}^{N}\hat{P}^{Z}\rangle}{\langle\hat{P}^{N}\hat{P}^{Z}\rangle}-\lambda_{q_{20}}\langle\hat{Q}_{20}\rangle-\lambda_{q_{22}}\langle\hat{Q}_{22}\rangle
\label{vap_equat}
\end{equation}
where the Lagrange multipliers, $\lambda_{q_{20}}$ and $\lambda_{q_{22}}$, guarantee the condition for the quadrupole moments, $\langle\hat{Q}_{20}\rangle=q_{20}$ and $\langle\hat{Q}_{22}\rangle=q_{22}$, with $q_{20}=\frac{\beta_{2}\cos\gamma}{C}$, $q_{22}=\frac{\beta_{2}\sin\gamma}{\sqrt{2}C}$ and $C=\sqrt{\frac{5}{4\pi}}\frac{4\pi}{3r_{0}^{2}A^{5/3}}$; $A$ is the mass number and $r_{0}=1.2$ fm.  

These intrinsic many-body states, $|\phi_{\beta_{2},\gamma}\rangle$, are subsequently projected to good number of protons and neutrons, and good angular momentum:
\begin{equation}
|JMK;NZ;\beta_{2},\gamma\rangle=\hat{P}^{J}_{MK}\hat{P}^{N}\hat{P}^{Z}|\phi_{\beta_{2},\gamma}\rangle
\label{proj_state}
\end{equation}
where $\hat{P}^{J}_{MK}$ is the angular momentum projector operator written in terms of an integral over the Euler angles~\cite{RingSchuck}.

Finally, the coefficients of the linear combination of Eq.~\ref{GCM_general}, $f^{J;NZ;\sigma}_{\lbrace\xi\rbrace}$ and the spectrum, $E^{J;NZ;\sigma}$, are obtained by solving the Hill-Wheeler-Griffin (HWG) equations~\cite{RingSchuck} that mix both quadrupole shapes and $K$, i.e., $\lbrace\xi\rbrace\equiv\lbrace\beta_{2},\gamma,K\rbrace$:
\begin{equation}
\sum_{\lbrace\xi'\rbrace}\left(\mathcal{H}_{\lbrace\xi\rbrace;\lbrace\xi'\rbrace}^{J;NZ}-E^{J;NZ;\sigma}\mathcal{N}_{\lbrace\xi\rbrace;\lbrace\xi'\rbrace}^{J;NZ}\right)f^{J;NZ;\sigma}_{\lbrace\xi\rbrace}=0
\label{HWG_eq}
\end{equation}
where $\mathcal{H}$ and $\mathcal{N}$ are the energy and norm overlaps respectively:
\begin{equation}
\mathcal{N}_{\lbrace\xi\rbrace,\lbrace\xi'\rbrace}^{J;NZ}=\langle JMK;NZ;\beta_{2},\gamma|JMK';NZ;\beta'_{2},\gamma'\rangle
\label{norm_kern}
\end{equation}
\begin{equation}
\mathcal{H}_{\lbrace\xi\rbrace,\lbrace\xi'\rbrace}^{J;NZ}=\langle JMK;NZ;\beta_{2},\gamma|\hat{H}|JMK';NZ;\beta'_{2},\gamma'\rangle
\label{ham_kern}
\end{equation} 
The generalized eigenvalue problem defined by Eq.~\ref{HWG_eq} for each value of the angular momentum is solved by transforming it into a regular eigenvalue equation in the following manner. First, the norm overlap matrix is diagonalized:
\begin{equation}
\sum_{\lbrace\xi'\rbrace}\mathcal{N}_{\lbrace\xi\rbrace,\lbrace\xi'\rbrace}^{J;NZ}u^{J;NZ}_{\lbrace\xi'\rbrace;\Lambda}=n^{J;NZ}_{\Lambda}u^{J;NZ}_{\lbrace\xi\rbrace;\Lambda}
\end{equation}
Then, an orthonormal set of states (the \textit{natural basis}) is obtained through the eigenvalues different from zero and their corresponding eigenvectors of the norm overlap matrix:
\begin{equation}
|\Lambda^{J;NZ}\rangle=\sum_{\lbrace\xi\rbrace}\frac{u^{J;NZ}_{\lbrace\xi\rbrace;\Lambda}}{\sqrt{n^{J;NZ}_{\Lambda}}}|JMK;NZ;\beta_{2},\gamma\rangle;\,n^{J;NZ}_{\Lambda}>0
\end{equation}
Finally, the original GCM state (Eq.~\ref{GCM_general}) can be written as:
\begin{equation}
|\Phi^{J;NZ;\sigma}\rangle=\sum_{\Lambda}G^{J;NZ;\sigma}_{\Lambda}|\Lambda^{J;NZ}\rangle
\end{equation}
and the HWG equations are transformed for each value of $J$ into a normal eigenvalue problem:
\begin{equation}
\sum_{\Lambda'}\langle\Lambda^{J;NZ}|\hat{H}|\Lambda'^{J;NZ}\rangle G^{J;NZ;\sigma}_{\Lambda'}=E^{J;NZ}G^{J;NZ;\sigma}_{\Lambda}
\label{HW}
\end{equation}
Expectation values and transition probabilities can be evaluated from the coefficients $G^{J;NZ}$ and the definition of the natural basis~\cite{RMP_75_121_2003,NPA_709_201_2002,PRC_78_024309_2008,PRC_81_064323_2010,PRC_81_044311_2010}. For example, the expectation value of a generic scalar operator under rotations, $O^{J;NZ;\sigma}\equiv\langle\Phi^{J;NZ;\sigma}|\hat{O}|\Phi^{J;NZ;\sigma}\rangle$, is computed within the GCM framework as:
\begin{widetext}
\begin{equation}
O^{J;NZ;\sigma}=\sum_{\Lambda\Lambda'}\sum_{\lbrace\xi\rbrace,\lbrace\xi'\rbrace}\left(G^{J;NZ}_{\Lambda}\frac{u^{J;NZ}_{\lbrace\xi\rbrace;\Lambda}}{\sqrt{n^{J;NZ}_{\Lambda}}}\right)^{*}\langle JMK;NZ;\beta_{2},\gamma|\hat{O}|JMK';NZ;\beta'_{2},\gamma'\rangle \left(G^{J;NZ}_{\Lambda'}\frac{u^{J;NZ}_{\lbrace\xi'\rbrace;\Lambda'}}{\sqrt{n^{J;NZ}_{\Lambda'}}}\right)
\label{GCM_EV}
\end{equation}
\end{widetext}
\section{Occupation numbers of spherical orbits}\label{OCC_NUM}
After discussing the general formalism to obtain expectation values within the SCCM framework, its application to compute occupation numbers of spherical orbits is sketched in this section. In this work the spherical orbits are obtained for each nucleus in a self-consistent manner from its spherically-symmetric HFB solution. By doing so, the arbitrariness in the definition of these non-observable quantities is partially removed. Hence, the operator associated to the number of particles occupying a given spherical orbit, $\alpha$, defined by the quantum numbers $(n_{\alpha}l_{\alpha}j_{\alpha})$ is:
\begin{equation}
\hat{n}_{\alpha}=\sum_{m_{j_{\alpha}}}a^{\dagger}_{n_{\alpha}l_{\alpha}j_{\alpha}m_{j_{\alpha}}}a_{n_{\alpha}l_{\alpha}j_{\alpha}m_{j_{\alpha}}}
\label{occ_num_op}
\end{equation}
These creation and annihilation single-particle operators $(a^{\dagger}_{\alpha},a_{\alpha})$ are obtained from the diagonalization of the density-matrix, $\rho^{sph}_{ab}$, that corresponds to the solution of a HFB calculation performed imposing spherical symmetry for the nucleus of interest. The HFB density matrix is defined as~\cite{RingSchuck}:
\begin{equation}
\rho^{sph}_{ab}=\langle\phi^{sph}|c^{\dagger}_{b}c_{a}|\phi^{sph}\rangle
\label{density_matrix_c}
\end{equation}
where $|\phi^{sph}\rangle$ is the spherical quasiparticle vacuum and $(c^{\dagger}_{a},c_{a})$ are creation and annihilation single-particle operators that correspond to the arbitrary working basis used to define the HFB transformation~\cite{RingSchuck}. This arbitrary basis is usually chosen to be a spherical harmonic oscillator basis made of a large number of major harmonic oscillator shells. By construction, the density-matrix expressed in terms of the single-particle operators $(a^{\dagger}_{\alpha},a_{\alpha})$ -usually known as \textit{canonical basis}- is diagonal:
\begin{equation}
\sum_{ab}A_{\alpha a}A^{*}_{\beta b}\rho^{sph}_{ab}=\tilde{\rho}_{\alpha\beta}=\langle\phi^{sph}|a^{\dagger}_{\beta}a_{\alpha}|\phi^{sph}\rangle=\tilde{\rho}_{\alpha\beta}\delta_{\alpha\beta}
\end{equation}

Obviously, working and canonical bases are related by the diagonalization matrix $A$, i.e., $a^{\dagger}_{\alpha}=\sum_{a}A_{\alpha a}c^{\dagger}_{a}$. Therefore, the one-body operator associated to the number of particles lying in a given spherical orbit can be expressed in a second quantization representation as: 
\begin{equation}
\hat{n}_{\alpha}=\sum_{ab}A_{\alpha a}A^{*}_{\alpha b}c^{\dagger}_{a}c_{b}
\label{ope_sq}
\end{equation}
Hence, the above expression can be used in Eq.~\ref{GCM_EV}, substituting $\hat{O}=\hat{n}_{\alpha}$ to evaluate the occupation numbers of the spherical orbits in the GCM -correlated- nuclear states. Additionally, the evolution of the occupation numbers with the intrinsic quadrupole deformation, $\beta_{2},\gamma$, is obtained from the diagonal part of the kernel given in Eq.~\ref{GCM_EV}, summing up the $K$-components:
\begin{equation}
n^{J;NZ}_{\alpha}(\beta_{2},\gamma)=\sum_{K}\frac{\langle JMK;NZ;\beta_{2}\gamma|\hat{n}_{\alpha}|JMK;NZ;\beta_{2},\gamma\rangle}{\langle JMK;NZ;\beta_{2},\gamma|JMK;NZ;\beta_{2},\gamma\rangle}
\label{occ_num_beta2}
\end{equation}

Finally, self-consistent single-particle energies (SPE) within the EDF framework can be defined in several ways, all of them related to the self-consistent one-body Hamiltonian matrix~\cite{RingSchuck}:
\begin{equation}
h_{pq}=t_{pq}+\Gamma_{pq}
\label{h_matrix}
\end{equation}
where $t_{pq}$ are the matrix elements of the one-body kinetic energy operator and $\Gamma_{pq}=\sum_{rs}\bar{v}_{prqs}\rho_{sr}$  is the Hartree-Fock field. Additionally, $\bar{v}_{prqs}$ are the antisymmetrized two-body matrix elements of the effective nuclear interaction and $\rho_{sr}$ the density matrix. The ambiguities in the definition of SPE come from the choice of the density matrix and the one-body basis in which such energies are evaluated. On the one hand, different density matrices can be obtained whether a HF or HFB calculation is chosen. On the other hand, SPE can be defined either as the eigenvalues of $h$ or the diagonal part of such a matrix expressed in a given basis, e.g., the canonical basis. Moreover, since HF or HFB calculations can be performed with constraints along different degrees of freedom, different density matrices can be obtained as a function of the deformation or any other intrinsic variable and Nilsson-like plots can be computed using this scheme. 

In the present work, HFB calculations are used to define the HF field in Eq.~\ref{h_matrix}. Furthermore, SPE are obtained as the diagonal part of $h$ written in the canonical basis and the spherical SPE (SSPE) are those found with $\tilde{\rho}^{sph}_{\alpha\alpha}$. 
The reader is referred to Ref.~\cite{PRC_85_034330_2012} (and references therein) for a detailed discussion on SPE and their relation to observables such as excitation energies of neighboring magic nuclei.
\section{Analysis of \texorpdfstring{$N=40$}{Lg} neutron rich isotones}\label{Results1}
\begin{figure}[t]
\begin{center}
\includegraphics[width=\columnwidth]{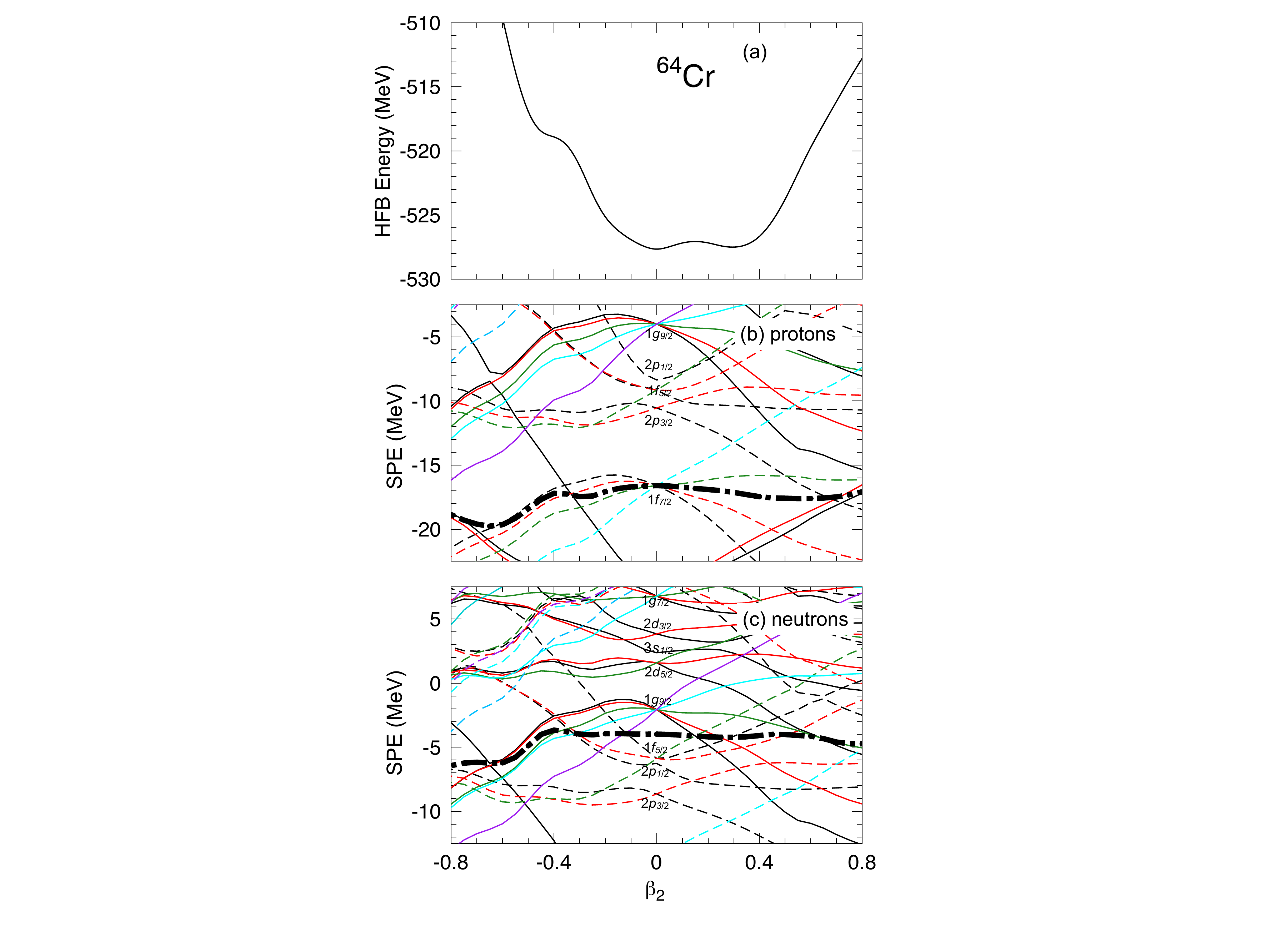}
\caption{(color online)  (a) HFB-Potential energy surface and HFB single-particle energies for (b) protons and (c) neutrons  as a function of the axial quadrupole deformation calculated for $^{64}$Cr with Gogny D1S interaction. 
\label{Fig1}}
\end{center}
\end{figure}

As a first application of the calculation of occupation numbers in spherical shells within a correlated EDF method, the role of the spherical orbits on the structure of $N=40$ neutron rich nuclei is discussed in this section. Recently, this region has been widely studied both experimental~\cite{PRL_88_092501_2002,PRC_76_034303_2007,PRC_81_051304_2010,PRC_81_061301_2010,PRC_84_061305_2011,PRL_106_022502_2011,PRC_86_011305_2012,PRC_86_014325_2012,PRL_110_242701_2013,PRC_87_014325_2013,PRC_88_024326_2013,PRC_88_041302_2013,PRC_89_021301_2013,PRL_112_112503_2014,PRC_91_034310_2015,PRC_92_034306_2015} and theoretically~\cite{EPJA_15_145_2002,PRC_78_064312_2008,PRC_80_064313_2009,PRC_81_051302_2010,PRC_82_031304_2010,PRC_82_054301_2010,PRC_83_061302_2011,PRC_86_024316_2012,PRC_87_047301_2013,PRC_89_024319_2014,PRC_90_054314_2014,PRC_91_024320_2015,JPG_42_045108_2015}  because of its interest as a new \textit{island of inversion} analogous to the one found at $N=20$. Here, a detailed analysis performed in the $^{64}$Cr nucleus is presented first and then, such a study is extended to the neutron rich $N=40$ even-even isotones, namely, $^{60}$Ca,  $^{62}$Ti, $^{64}$Cr, $^{66}$Fe and $^{68}$Ni. Calculations are performed with the Gogny D1S parametrization~\cite{Berger84} using a working basis made of eleven major spherical harmonic oscillator shells. In addition, the calculations are simplified and only axial-symmetric intrinsic HFB states (spherical $(\beta_{2}=0)$, prolate $(\gamma=0^{\circ},\beta_{2}>0)$ and oblate $(\gamma=180^{\circ},\beta_{2}<0)$) have been considered in this work. Therefore, the expressions given above are also reduced to $K=0$ components and the quadrupole deformation $\beta_{2}$ is the only remaining generating coordinate. 
\subsection{Occupation numbers for \texorpdfstring{$^{64}$}{Lg}Cr}
From a self-consistent mean-field point of view the usual starting point to describe the structure of a given nucleus is the calculation of the mean-field energy as a function of the most relevant collective coordinates like the quadrupole deformation. In Fig.~\ref{Fig1} a constrained HFB calculation of the nucleus $^{64}$Cr is shown as an example. In fact, HFB equations are a simplified version of Eq.~\ref{vap_equat} where the particle number projectors are set to the identity. Hence, the HFB energy as a function of the axial quadrupole deformation, i.e., the potential energy surface (PES) is shown in Fig.~\ref{Fig1}(a). Two almost degenerated minima are obtained, the absolute minimum in the spherical point and another one at a prolate deformation $\beta_{2}=0.35$. Furthermore, SPE close to the Fermi energies (plotted as thick dot-dashed lines) are shown in Fig.~\ref{Fig1}(b)-(c) for protons and neutron respectively. Spherical orbits with well-defined ($n,l,j$) quantum numbers and $(2j+1)$ degeneracies, i.e., SSPE, are obtained at $\beta_{2}=0$. Such a degeneracy is broken when the deformation is increased and Nilsson-like orbits are obtained. Normally, the minima found in the PES can be related to the appearance of sizable gaps in the SPE crossed by the Fermi level. In the present example, the neutron Fermi energy crosses the gap between the $1f_{5/2}$ and $1g_{9/2}$ orbits, producing the spherical minimum. Moreover, the prolate minimum can be related to the gap produced by the lowering of some levels coming from the neutron $1g_{9/2}$ and the rising of levels coming from the neutron $1f_{5/2}$ orbits, in combination with the gap found in the proton SPE due to the breaking of the spherical degeneracy of the $1f_{7/2}$ orbit.   
\begin{figure}[t]
\begin{center}
\includegraphics[width=\columnwidth]{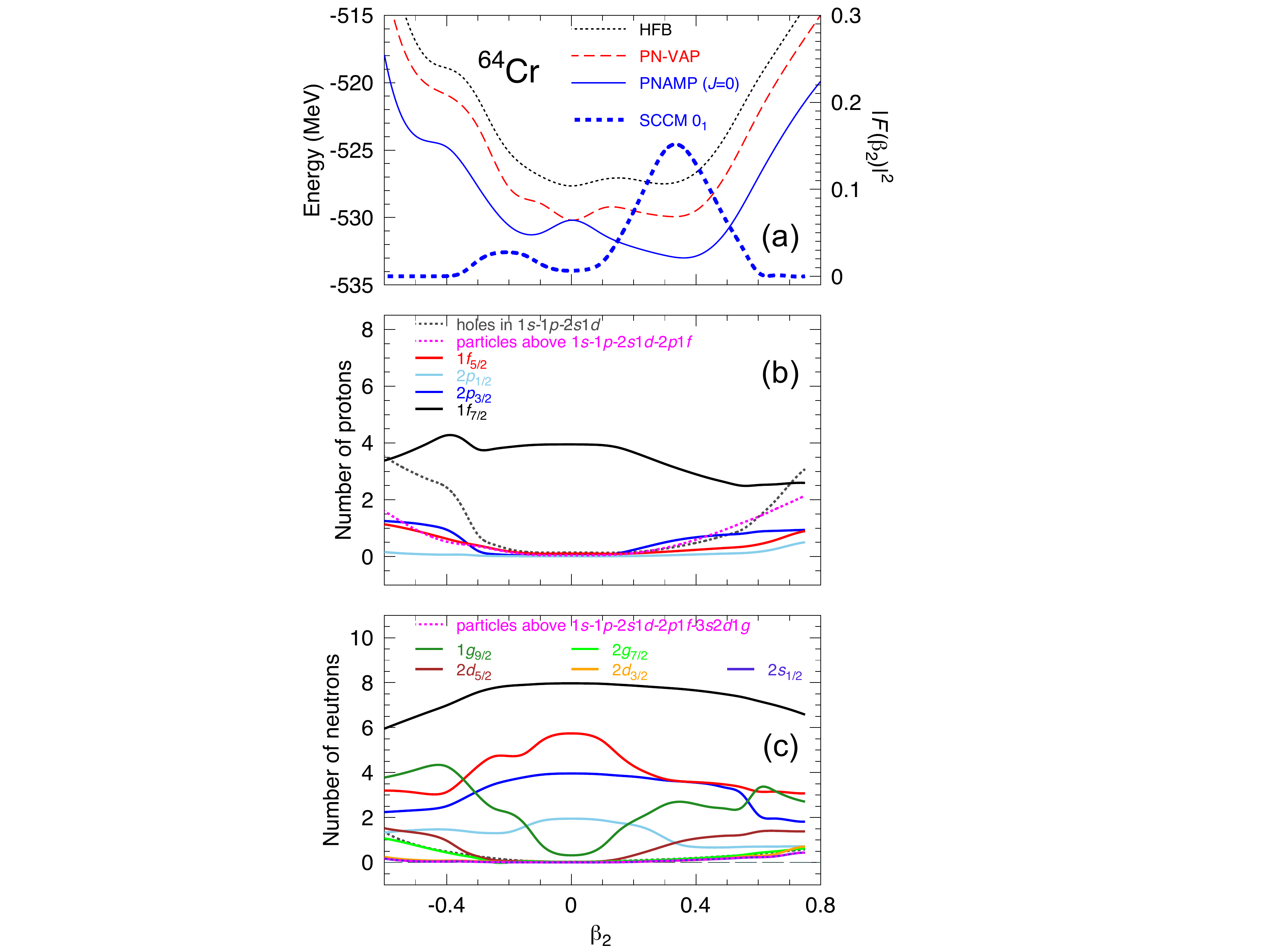}
\caption{(color online)  (a) Potential energy surface as a function of the axial quadrupole deformation calculated for the nucleus $^{64}$Cr with HFB (thin-dotted), PN-VAP (dashed) and particle-number and angular momentum, $J=0$, projection (continuous) approaches. In thick-dash line, the ground state collective wave-function is shown. (b) and (c) occupation numbers of spherical orbits as a function of the axial quadrupole deformation for protons and neutrons respectively.
\label{Fig2}}
\end{center}
\end{figure}
\begin{figure*}[t]
\begin{center}
\includegraphics[width=1.0\textwidth]{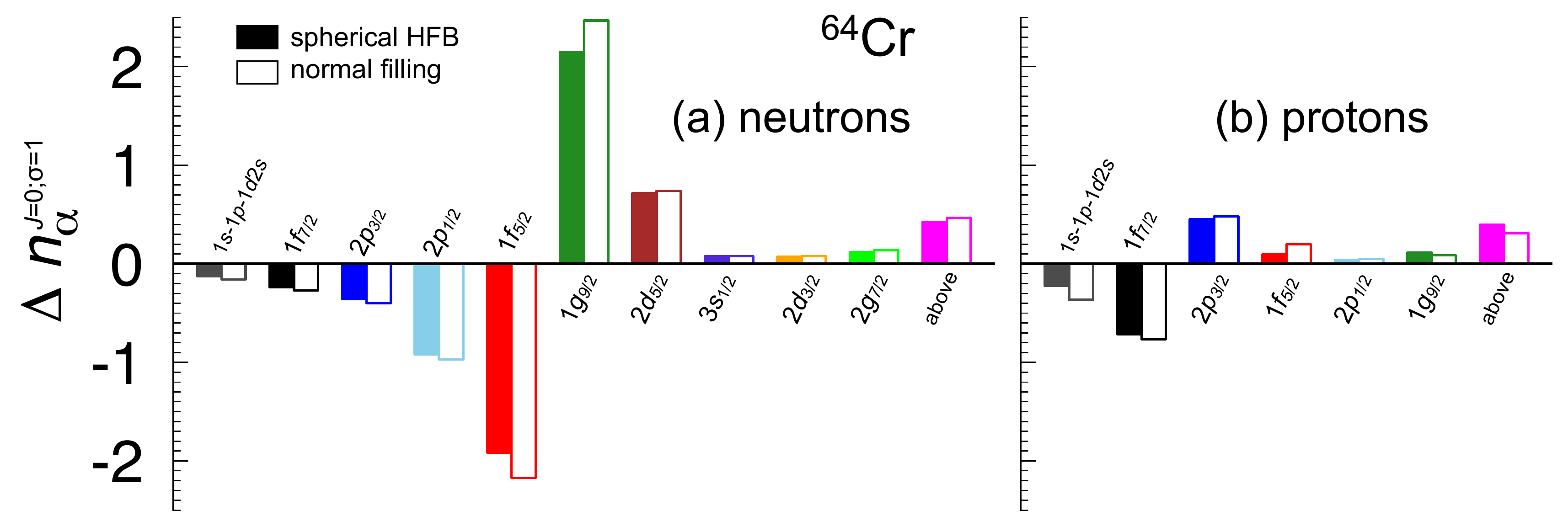}
\caption{(color online) Difference between the occupation numbers of spherical orbits for the ground state of $^{64}$Cr calculated with the SCCM method and the occupation numbers of those orbits in the spherical HFB configuration (filled bars) and in the normal filling approximation (empty bars).
\label{Fig3}}
\end{center}
\end{figure*}

At this level of approximation, the relevance of certain spherical single particle levels in the structure of the nucleus is only qualitatively established by their relative position to the Fermi energy. Hence, it is clear from Fig.~\ref{Fig1} that if the nucleus $^{64}$Cr is prolate deformed the neutron $1g_{9/2}$ ($1f_{5/2}$) orbit will be partially occupied (empty) contrary to what it is expected if the nucleus is spherical. However, a more quantitative description and within more correlated states are required to compare with SM results. In Fig.~\ref{Fig2}(a) the result of including first particle number restoration (PN-VAP) and particle number and angular momentum projection (PNAMP, $J=0$) is shown together with the HFB PES already mentioned. Here, the PN-VAP energy curve is similar to the HFB one but shifted to lower values. Additional correlation energy is obtained by performing the angular momentum projection. However, the absolute minimum in this case is the prolate one ($\beta_{2}=0.4$) and a secondary minimum at an oblate shape ($\beta_{2}=-0.15$) is also found, being the spherical point a maximum due to the impossibility of gaining rotational correlation energy with this specific shape. In fact, the prolate minimum is now around 3 MeV lower than the spherical configuration, showing the relevance of performing the angular momentum projection in nuclei where spherical and deformed shapes are competing at the mean-field level. Similar results are also found in $^{32}$Mg~\cite{NPA_709_201_2002} and $^{80}$Zr~\cite{PLB_705_255_2011}.  

Using these PNAMP states with $J=0$, $|J;N,Z;\beta_{2}\rangle$, and the spherical HFB state computed for $^{64}$Cr as the reference state to define the spherical orbits, the occupancies of these orbits as a function of the axial quadrupole deformation (Eq.~\ref{occ_num_beta2}) are plotted in Fig.~\ref{Fig2}(b)-(c) for protons and neutrons respectively. These figures are better understood in combination with Fig.~\ref{Fig1}(b)-(c). Focusing on the spherical point, $\beta_{2}=0$, practically the normal filling of protons shells is observed, i.e., protons ($Z=24$) occupy the first three major shells and the remaining 4 protons are in the $1f_{7/2}$ orbit. For neutrons ($N=40$), the lowest three major shells, $1f_{7/2}$, $2p_{3/2}$ and $2p_{1/2}$ are also filled in and, due to some remaining pairing correlations provided by the PN-VAP approach, the $1f_{5/2}$ orbit is not totally full but some occupancy is transferred to the $1g_{9/2}$ orbit. Differences between the normal filling (spherical HF) and the occupancies of the spherical levels in the HFB state with $\beta_{2}=0$ will be further discussed below.  

When the prolate deformation increases, the occupancy of the proton $1f_{7/2}$ slowly decreases in the interval $\beta_{2}\in\left[0.15,0.5\right]$ and, at the same time, some occupancies appear in the $2p_{3/2}$, $1f_{5/2}$ and above the $pf$-shell (mainly from the proton $1g_{9/2}$) and some vacancies from the $sd$-shell (mainly from the $1d_{3/2}$). The relevance of protons (holes) above (below) the $pf$-shell becomes important at large prolate deformations and also from $\beta_{2}=-0.3$ and larger oblate deformations. In the oblate part, similar behavior as in the prolate part of the proton $2p_{3/2}$, $1f_{5/2}$ orbits is obtained but the occupancy of the proton $1f_{7/2}$ shows a maximum at $\beta_{2}=-0.4$ due to the promotion of particles from the $2s_{1/2}$ orbit to this level.  

Concerning neutron occupation numbers, particles above (holes below) the $1g_{9/2}$ ($1f_{7/2}$) are rather small. Furthermore, the $1f_{7/2}$ and $2p_{3/2}$ orbits remain almost full in a wide range of deformation $\beta_{2}\in\left[-0.3,0.5\right]$. However, the number of neutrons in the $1g_{9/2}$ orbit increases as soon as the nucleus starts to be deformed due to the decrease in the occupancy of the $1f_{5/2}$ level. In fact, for $|\beta_{2}|>0.4$ the $1g_{9/2}$ level reaches an occupancy of around three neutrons. Furthermore, the $2p_{1/2}$ orbit empties and the $2d_{5/2}$ orbit fills in as soon as the deformation increases from the spherical point.    

Having discussed the occupancies of the spherical shells as a function of the quadrupole deformation, the final step consists in computing the occupation numbers taking into account the mixing of different shapes. First, the relevant deformations in the final states are given by the so-called collective wave function~\cite{RingSchuck,RMP_75_121_2003}, $|F(\beta_{2})|^{2}$, that represent the weights of the different intrinsic deformations (or collective coordinates) in a given nuclear state. These states are obtained after performing particle number and angular momentum projections and shape mixing within the SCCM framework described in Sec.~\ref{SCCM}. In Fig.~\ref{Fig2}(a) the ground-state collective wave function of the nucleus $^{64}$Cr is plotted with a thick dashed line. Here, the largest contributions correspond to prolate deformations with an absolute maximum at $\beta_{2}=0.35$. A secondary peak is found at oblate deformations ($\beta_{2}=-0.2$) although the contribution of such configurations is much smaller. These two peaks appear consistently at the position of the minima found in the PNAMP-PES. Therefore, although this nucleus is found to be spherical at the mean-field (HFB) level due to the $N=40$ harmonic oscillator shell closure, BMF correlations favor a prolate deformed ground state.

The basic results of our work for $^{64}$Cr are gathered in
Fig.~\ref{Fig3} where we plot first the difference between the occupation numbers of the most relevant spherical shells computed for the SCCM ground-state of $^{64}$Cr ($J=0;\sigma=1$), and those calculated for its spherical HFB wave function,  Such a difference is defined for any SCCM state as (see Eqs.~\ref{ope_sq},~\ref{GCM_general} and~\ref{GCM_EV}):
\begin{equation}
\Delta n^{J;\sigma}_{\alpha}=\langle\Phi^{J;\sigma}|\hat{n}_{\alpha}|\Phi^{J;\sigma}\rangle-\langle\phi^{sph}|\hat{n}_{\alpha}|\phi^{sph}\rangle
\end{equation}
Positive (negative) values of $\Delta n^{J;\sigma}_{\alpha}$ mean particles (holes) in a given SSP level $\alpha$ with respect to the filling in the spherical HFB configuration (filled bars).  In addition, we plot the differences between the final
SCCM occupancies and the normal filling given by the spherical HF solution (empty bars). The figure contains a lot of physical
information which we shall analyze in what follows.

For neutrons ($N=40$), Fig.~\ref{Fig3}(a), the $pf$ and $sdg$ shells are explicitly plotted while for protons ($Z=24$), Fig.~\ref{Fig3}(b), only the $pf$ shell and the $1g_{9/2}$ orbit are singled out. Looking at the differences with the spherical HFB configuration, the orbits below $1f_{7/2}$ are almost fully occupied although $\sim0.12$ ($\sim0.22$) neutron holes (proton holes) are obtained. Furthermore, above the neutron $sdg$ shell (proton $1g_{9/2}$ orbit)  around $\sim0.42$ ($\sim0.40$) particle excess is found. The main differences with respect to the spherical HFB solution are found in the depopulation of the neutron $pf$ shell and the occupation of the neutron $1g_{9/2}$ ($\sim 2.15$ particles) and $2d_{5/2}$ ($\sim 0.72$ particles) orbits, mainly. Most of the depopulation of the neutron $pf$ shell comes from the $1f_{5/2}$ ($\sim1.91$ holes) and $2p_{1/2}$ ($\sim0.91$ holes) levels and, to a lesser extent, the $2p_{3/2}$ ($\sim0.36$ holes) and $1f_{7/2}$ ($\sim0.24$ holes) orbits. It is also interesting to see that  $3s_{1/2}$, $2d_{3/2}$ and $2g_{7/2}$ are not very much populated in the ground state. This shows that the neutron valence space used in recent SM-CI calculations~\cite{PRC_82_054301_2010} to describe the onset of deformation in this region is supported by these results. Furthermore, they emphasize the importance of including in the valence space not only the $1g_{9/2}$ orbit but also its quasi-SU(3) partner $2d_{5/2}$.  In fact, the pseudo+quasi SU(3) model of Ref. \cite{PRC_92_024320_2015} predicts for $^{64}$Cr a dominant 4p-4h neutron configuration, with
2.3, 1.5 and 0.2 neutrons in the $1g_{9/2}$,  $2d_{5/2}$, and  $3s_{1/2}$ orbits respectively. 
Deformation also influences the proton occupancies as it is shown in Fig.~\ref{Fig3}(b). Here, the $1f_{7/2}$ orbit does not contain anymore the four valence protons as in the spherical case but it accommodates roughly one proton less, while the $2p_{3/2}$ and $2f_{5/2}$ starts to be slightly occupied. 
The occupancies of the SM-CI calculation in the LNPS valence space and the values obtained in the quasi+pseudo SU(3) model 
are compared with the SCCM, spherical HFB and spherical HF (uniform filling) values in table \ref{tab:cr64occ}. 

Before making the detailed comparison of the SCCM and SM-CI results, let us highlight some of the findings of Fig.~\ref{Fig3}. The first one concerns the role of pairing and stems from the comparison of the HF and HFB occupancies
(see also  table~\ref{tab:cr64occ}). It is seen that the effect of the pairing interaction is limited to a few orbits above
and below the Fermi level, what we can dub, the "natural" shell model valence space. And even so, the number of scattered pairs is quite small, due to the presence of large gaps in neutrons and protons associated to the $N$=40 and $Z$=28 magic numbers. It is only when the quadrupole correlations are duly taken into account that deformation sets in, blowing out these shell closures. Another interesting feature relates to the 2$\hbar \omega$ excitations which drive the coupling to the GDR. Notice that the vacancies of neutrons below $N=28$ and protons below $Z=20$, the real core of the SM-CI calculation, amount only to 0.5 each. However, any calculation excluding them would need to use effective charges (or masses) to reproduce the experimental data for E2 transitions and spectroscopic quadrupole moments. The analysis of Ref.~\cite{PRC_54_1641_1996} explains how these perturbative effects produce the standard isoscalar effective charge $\delta q_{\pi} + \delta q_{\nu}$=0.77.

\begin{table}[b]
\begin{tabular*}{\linewidth}{@{\extracolsep{\fill}}|c|ccccccccc|}
\hline  
  & (b) & $1f^{\nu}_{7/2}$ & $2p^{\nu}_{3/2}$ & $2p^{\nu}_{1/2}$ & $1f^{\nu}_{5/2}$ & $1g^{\nu}_{9/2}$ & $2d^{\nu}_{5/2}$ & $3s^{\nu}_{1/2}$& (a) \\
\hline
HF$_{\mathrm{sph}}$ & 20.0 & 8.0 & 4.0 & 2.0 & 6.0 & 0.0 & 0.0 & 0.0 & 0.0 \\
HFB$_{\mathrm{sph}}$ & 20.0 & 8.0 & 4.0 & 1.9 & 5.7 & 0.3 & 0.0 & 0.0 & 0.1 \\
SCCM & 19.8 & 7.7 & 3.4 & 1.0 & 3.7 & 2.6 & 0.9 & 0.1 & 0.8 \\
SM-CI & 20.0 & 8.0  & 3.9  & 1.0  & 3.2  & 3.2 &0.7  & 0.0 & 0.0 \\
SU(3) & 20.0 & 8.0 & 2.8 & 1.4 & 3.8 & 2.3 & 1.5 & 0.2 & 0.0 \\
\hline 
  & (b) & $1f^{\pi}_{7/2}$ & $2p^{\pi}_{3/2}$ & $2p^{\pi}_{1/2}$ & $1f^{\pi}_{5/2}$ & $1g^{\pi}_{9/2}$ & (a) &  & \\ 
\hline 
HF$_{\mathrm{sph}}$& 20.0 & 4.0 & 0.0 & 0.0 & 0.0 & 0.0 & 0.0 &  &  \\
HFB$_{\mathrm{sph}}$& 19.9 & 4.0 & 0.0 & 0.0 & 0.1 & 0.0 & 0.0 & & \\
SCCM  & 19.5 & 3.2 & 0.6 & 0.1 & 0.2 & 0.1 & 0.3 &  &  \\
SM-CI & 20.0 & 3.1 & 0.4 & 0.2& 0.3  & 0.0  & 0.0 &  & \\
SU(3) & 20.0 & 2.7 & 1.3 & 0.0 & 0.0  & 0.0 & 0.0 &  & \\
 \hline 
\end{tabular*}
\caption{\label{tab:cr64occ} Occupation numbers of the spherical orbits in the ground state of $^{64}$Cr, for the different approaches discussed in the text. (b) and (a) refer to particles below and  above the orbits explicitly shown.}
\end{table}  
Back to Table \ref{tab:cr64occ}, we can make these statements more quantitative. In the neutron side, the occupancies predicted by the SCCM and SM-CI calculations are astonishingly similar. Notice as well that both calculations resemble qualitatively to the values obtained in the quasi+pseudo SU(3) limit. The agreement is even better in the proton sector. All in all we can conclude that, once the deformed regime is established, the dominance of the quadrupole-quadrupole interaction is well described by either SCCM or SM-CI, whereas, the most relevant physics features can be already captured by an algebraic model based in variants of Elliott's SU(3).
\begin{figure}[t]
\begin{center}
\includegraphics[width=\columnwidth]{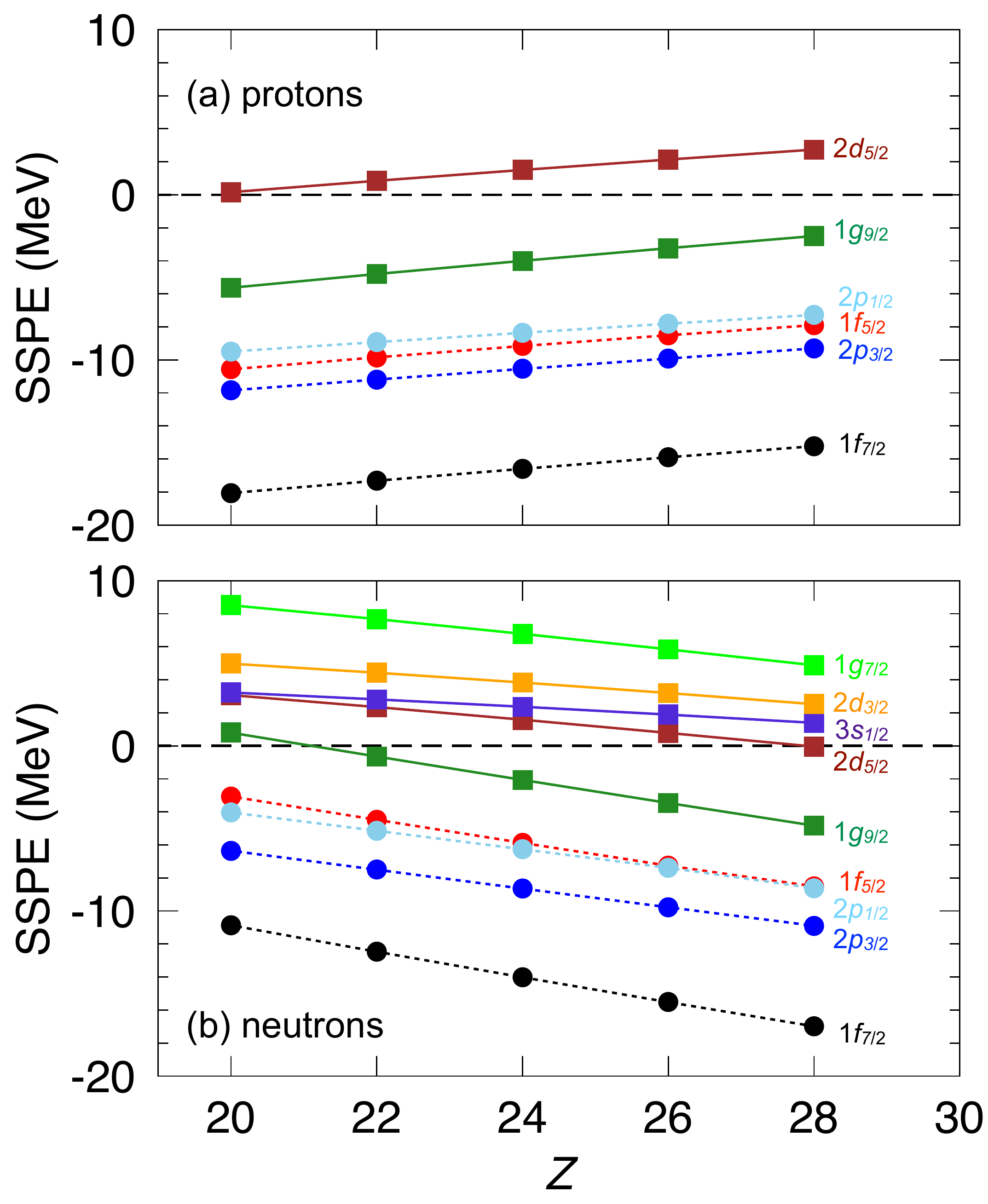}
\caption{(color online)  HFB-spherical single-particle energies -(a) proton and (b) neutron levels, respectively- calculated for neutron rich $N=40$ isotones with Gogny D1S interaction. 
\label{Fig4}}
\end{center}
\end{figure}
\subsection{Occupation numbers and deformation for the \texorpdfstring{$N=40$}{Lg} neutron rich isotones.}
The analysis that we have carried out for $^{64}$Cr can be extended to the $N=40$ even-even nuclei, from $^{60}$Ca to $^{68}$Ni. The appearance of deformation is usually discussed within the SM-CI framework in terms of the competition between the spherical mean-field gaps and the quadrupole correlations. Hence, if the effective single-particle energies (ESPE) of the intruder levels get closer to those of the lowest level occupied in a normal filling approach, the quadrupole interactions can favor energetically deformed $n$-particle $n$-hole configurations across the gap (intruder states). The same qualitative study can be performed by representing the spherical single-particle energies (SSPE) obtained with HFB spherical states that have been computed with the Gogny D1S interaction (see Sec.~\ref{SCCM}). 

In Fig.~\ref{Fig4} the most relevant SSPE's are plotted both for protons and neutrons in the range $20\leq Z\leq 28$ and $N=40$ (see also Ref.~\cite{PRC_80_064313_2009}). In the proton sector, three large gaps are observed, namely: a) the $Z=28$ gap ($\sim6$ MeV) between the $1f_{7/2}$ orbit and the $2p_{3/2}$, $1f_{5/2}$ and $2p_{1/2}$ pseudo-spin triplet; b) the $Z=40$ gap ($\sim4$ MeV) between the $2p_{1/2}$ and $1g_{9/2}$ orbits; and c) the $Z=50$ gap ($\sim6$ MeV) between the $1g_{9/2}$ and the $2d_{5/2}$ orbits. These gaps are rather constant along the isotonic chain and no erosion of the spherical harmonic oscillator plus spin-orbit shell closures is found as the neutron excess increases.

Concerning the neutron SSPE's (Fig.~\ref{Fig4}(b)), the spherical harmonic oscillator plus spin orbit gaps are also clearly observed. However, although the $N=28$, 40 and 50 gaps are rather robust along the isotonic chain, these are not as constant as the proton ones. On the one hand, the gap between the $1f_{7/2}$ and $2p_{3/2}$ orbits tends to get reduced in neutron rich nuclei (from $\sim6$ MeV to $\sim4.5$ MeV). Similarly, the $N=50$ gap is significantly reduced in $^{60}$Ca with respect to the one obtained for $^{68}$Ni (from $\sim5$ MeV to $\sim2$ MeV). Furthermore, a small gap ($\sim2$ MeV) between the $2p_{3/2}$ and the $2p_{1/2}$-$1f_{5/2}$ levels ($N=32$) is obtained. The gap between the latter and the $1g_{9/2}$ orbit remains almost constant ($\sim3.5$ MeV). Finally, the $2d_{5/2}$ and $3s_{1/2}$ orbits are almost degenerated in $^{60}$Ca but they split apart slightly when more protons are added into the system.

This picture differs somehow from the ESPE obtained with LSSM calculations~\cite{PRC_82_054301_2010}. Whereas the proton gaps for all $Z$ values and the neutron gaps in $^{68}$Ni are quite similar in both approaches, the evolution of the neutron gaps towards $^{60}$Ca is very different. The LNPS interaction predicts that the orbits $1f_{5/2}$, $1g_{9/2}$ and $2d_{5/2}$ become degenerated in the neutron rich part ($^{60}$Ca), while with the Gogny SSPE's the $N$=40 gap remains constant. This makes the two approaches to diverge in their predictions of the structure of  $^{62}$Ti and $^{60}$Ca, as we shall discuss below. The degeneracy of the ESPE's favors the persistence of quadrupole correlations, responsible for the onset of deformation. These differences will manifest clearly in the occupation numbers and in the quadrupole deformation parameters produced by both descriptions. 
Let us add that there have been recent Coupled Cluster calculations around $^{60}$Ca, using Chiral EFT and including in an effective way three body forces and the effect of the continuum (see Ref. \cite{prl_109_032502_2012}). Their conclusions are midway between the two approaches discussed here. They find that the ordering of levels in $^{61}$Ca is inverted with respect to the standard shell model filling with a sequence 1/2$^+$, 5/2$^+$, 9/2$^+$, closer to the SM-CI ESPE's. However, they propose a configuration $(3s_{1/2})^{2\nu}$ for the ground state of $^{62}$Ca, which implies a certain resilience  of the $N=40$ gap, as suggested also by the SCCM description. Indeed, only spectroscopic data on $^{62}$Ti settle these discrepancies.

\begin{figure}[t]
\begin{center}
\includegraphics[width=\columnwidth]{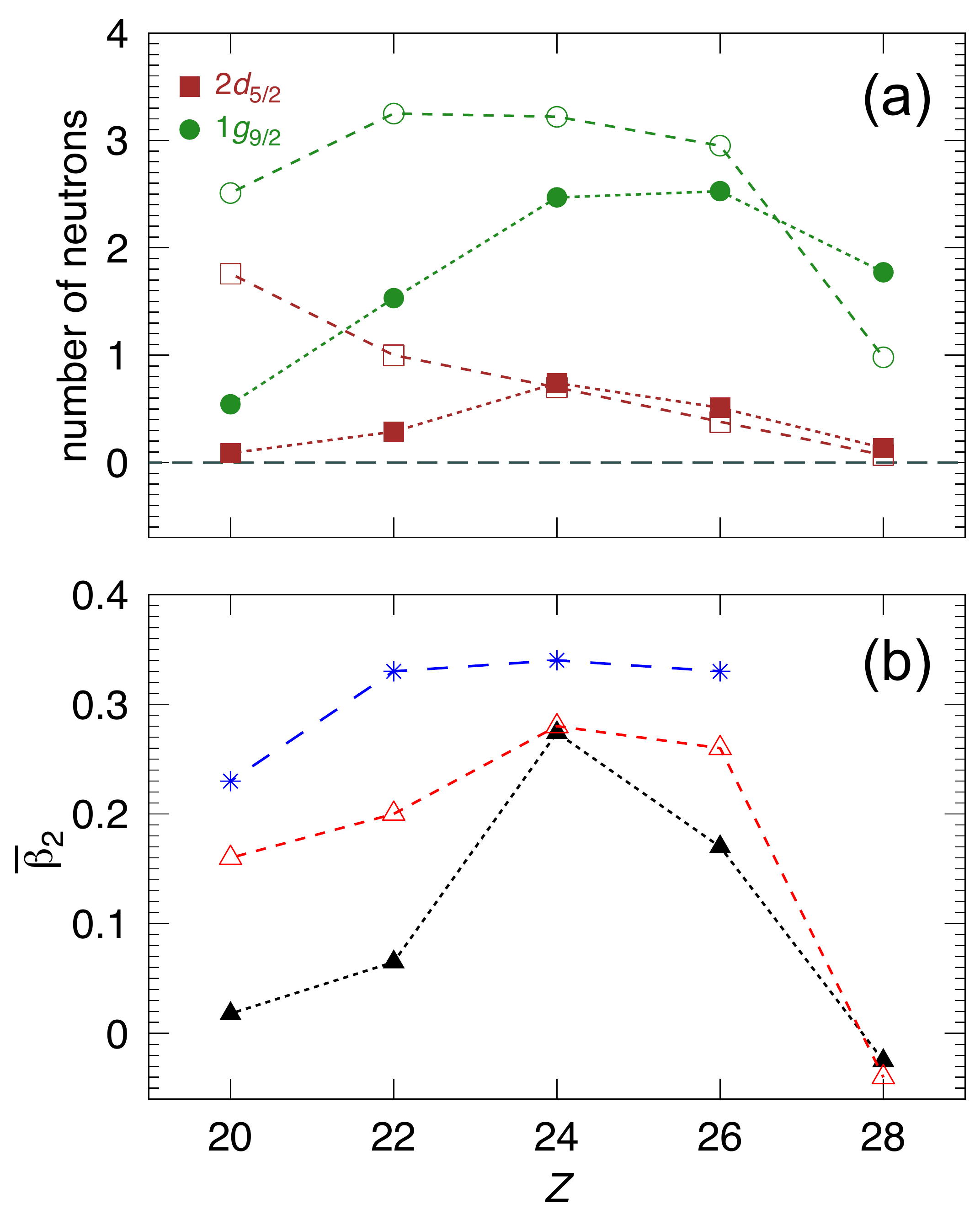}
\caption{(color online) (a) Occupation numbers of the neutron $1g_{9/2}$ and $2d_{5/2}$ spherical orbits for the ground states of the $N=40$ isotones. (b) Mean-values of the quadrupole deformation for the same states. Filled and open symbols refer to SCCM and SM-CI results respectively. Asterisks represents the quasi+pseudo SU(3) regime.
\label{Fig5}}
\end{center}
\end{figure}

The occupation numbers of the $1g_{9/2}$ and $2d_{5/2}$ spherical neutron orbits are plotted in Fig.~\ref{Fig5}(a). Indeed, the differences in the ESPE's reflect directly in the values shown in the figure. The agreement that we have found in the case of $^{64}$Cr extends to $^{66}$Fe and to a lesser extent to $^{68}$Ni. Notice that the $2d_{5/2}$ orbit has a non-negligible occupancy only when the nucleus is deformed, hence the large differences between SCCM and SM-CI for $^{62}$Ti and $^{60}$Ca.
In the SCCM approach, the $2d_{5/2}$ and $3s_{1/2}$ (not shown) orbits present a similar behavior, i.e., their occupation is maximum in the middle of the chain ($Z=24$) and negligible at the shell closures. Nevertheless, the $3s_{1/2}$ orbit is almost empty and this fact justifies its exclusion from the valence space in LSSM calculations~\cite{PRC_82_054301_2010}. There is a clear correlation between the deformation parameter and the simultaneous occupation of both the $1g_{9/2}$ \textit{and} $2d_{5/2}$ neutron orbits, which we proceed to discuss in some detail.

The ground-state (mass) quadrupole deformation can be computed within the SCCM method by taking the value of $\beta_{2}$ weighted by the ground-state collective wave function. In Fig.~\ref{Fig5}(b) such a mean-value, $\bar{\beta}_{2}$, is represented for the isotonic chain $N=40$. Here, almost spherical shapes are obtained at the proton magic numbers, namely, $^{60}$Ca ($\beta_{2}=+0.02$) and $^{68}$Ni ($\beta_{2}=-0.03$). Furthermore, the most deformed nucleus in the isotonic chain is found at $Z=24$ ($^{64}$Cr) with a prolate deformation equal to $\beta_{2}=+0.27$. In the SM-CI description, we compute the spectroscopic quadrupole moments and the $B(E2)$ transitions of the yrast band, with the standard isoscalar effective "mass". In the ideal rotor, the intrinsic quadrupole moments extracted from any of these observables using the well known Bohr-Mottelson formulas~\cite{BM_book} should be all the same. When deviations occur, we average the different values.
The $B(E2)$ values are known experimentally for $^{64}$Cr and $^{66}$Fe (see for instance reference \cite{PRL_110_242701_2013}) and the SM-CI  calculation reproduces them perfectly. Therefore, the SM-CI points in Fig.~\ref{Fig5}(b) may serve as experimental data as well. Notice again that the agreement between SCCM and SM-CI for these two nuclei is very good. The accord is excellent for $^{68}$Ni as well, in spite of the discrepancy in the filling of the orbit $1g_{9/2}$. To extract the ground state deformation of a quasi spherical nucleus in the SM-CI context can be tricky (or even nonsensical). We have proceeded as follows; the first excited 2$^+$ state is mildly oblate and makes a kind of band with the first excited 0$^+$ state. Using its spectroscopic quadrupole moment and the ratio of the $B(E2)$'s to the two  0$^+$ states, we estimate the amount of mixing of spherical and oblate components in the ground state and then compute the average deformation. In Fig.~\ref{Fig5}(b) we have plotted the deformation parameters computed in the quasi+psudo SU(3) limit, which can be taken as upper bounds for the real ones. The SM-CI values follow the trend of the SU(3) prediction with typically a 20\% reduction. In  $^{62}$Ti and $^{60}$Ca the SCCM deformation parameters depart drastically of this limit, as anticipated in view of the occupancies of the spherical orbits. $^{62}$Ti is the key nucleus to settle the evolution of the ESPE in this region.  

Finally let us examine closely the case in which the two approaches diverge the most, $^{60}$Ca. As mentioned before, the SCCM method makes it spherical and doubly magic, as seen both in the occupancies and in the deformation parameter. But, what is the SM-CI image of this nucleus?. Common lore associates deformation to the presence of neutrons and protons in open orbits. Indeed, in the present SM-CI calculation there are no active protons at all; the often used parameter N$_p$N$_n$ is just null. Neutrons alone in degenerate orbits seem to call for some kind of superfluid regime, but we shall show that this not the case at all. In the quasi+pseudo SU(3) regime, $^{60}$Ca has an yrast band with a perfect $J(J+1)$ spacing and constant values of the intrinsic mass quadrupole moment $Q^m_0$= 130~fm$^2$ or $\beta^m_2$= 0.23. Surprisingly, the calculation with the realistic interaction LNPS produces results that are much closer to the quadrupole than to the pairing limit.
The  yrast energies are distorted by the pairing interaction and depart from the $J(J+1)$ law, with $E(4^+)/E(2^+)=2.2$ (in the pairing limit this ratio is equal to 1). However, from the E2 observables we can extract a value $Q^m_0=100(5)$~fm$^2$ or $\beta^m_2$= 0.18(1), consistent with a deformed rotor. This is an unexpected fact (perhaps only of academic value), which shows that, if the single particle orbits around the Fermi surface map the SU(3) favoring quantum numbers, and if they are quasi degenerated, deformation may set in, even in the case that only alike particles are active in the natural valence space.
\section{Summary}\label{Summary_outlook}
In this article we develop a method to compute occupation numbers of spherical orbits within an energy density functional framework  based on the Gogny interaction that includes beyond-mean-field effects (symmetry restorations and quadrupole shape mixing). The nuclear states are computed with a symmetry conserving configuration mixing method and then used to calculate the expectation values of the operators that define the spherical orbits. These are determined self-consistently for each nucleus as the canonical basis of a spherical Hartree-Fock-Bogoliubov calculation.
As a first application of the method, the single-particle structure of the ground state of the nucleus $^{64}$Cr has been studied. This analysis has been extended to other neutron rich $N=40$ isotones from $Z=20-28$, showing the role of the neutron $1g_{9/2}$ and $2d_{5/2}$ orbits in the onset of deformation in this region. 

All these results are compared with "state-of-the-art" large scale shell model calculations (SM-CI). Since the SCCM method does not have an inert  core,  uses a very large number of major harmonic oscillator shells, and the underlying interaction is of general applicabiliy, the evaluation of the spherical occupation numbers can be done everywhere in the nuclear chart. Such information, a) can provide a better understanding of the single-particle structure of the nuclear states obtained with SCCM calculations; b) can be compare directly with LSSM results; and c) can help defining physically sound valence spaces for LSSM calculations.   
In the near future, the calculation of the number of nucleons occupying spherical shells will be extended to include other relevant degrees of freedom such as octupolarity, triaxiality and/or time-reversal symmetry breaking in the intrinsic wave functions.

\section*{Acknowledgements}
We thank G. Mart\'inez-Pinedo and L. M. Robledo for helpful discussions. We acknowledge the support from GSI-Darmstadt and CSC-Loewe-Frankfurt computing facilities; from the Ministerio de Econom\'ia y Competitividad (Spain) under contracts FIS2014-53434, FPA2014-57196, and Programas Ram\'on y Cajal 2012 number 11420 and 
Centros de Excelencia Severo Ochoa SEV-2012-0249; and from the USIAS program
of the University of Strasbourg.


\end{document}